\documentclass[twocolumn,aps,prl,showpacs,superscriptaddress,floatfix]{revtex4}
\pdfoutput=1
\usepackage{graphicx}

\begin{document}

\title{Phase Coherence and Superfluid-Insulator Transition in a Disordered Bose-Einstein Condensate}

\author{Yong P. Chen}
\altaffiliation[Current Address: ]{Purdue University, 525 Northwestern Ave., West Lafayette, IN 47907, USA}
\affiliation{Department of Physics and Astronomy and Rice Quantum Institute, Rice University, 6100 Main St., Houston TX 77005, USA}
\affiliation{The Richard E. Smalley Institute for Nanoscale Science and Technology, Rice University, 6100 Main St., Houston TX 77005, USA}
\author{J.~Hitchcock}
\affiliation{Department of Physics and Astronomy and Rice Quantum Institute, Rice University, 6100 Main St., Houston TX 77005, USA}
\author{D.~Dries}
\affiliation{Department of Physics and Astronomy and Rice Quantum Institute, Rice University, 6100 Main St., Houston TX 77005, USA}
\author{M.~Junker}
\affiliation{Department of Physics and Astronomy and Rice Quantum Institute, Rice University, 6100 Main St., Houston TX 77005, USA}
\author{C.~Welford}
\affiliation{Department of Physics and Astronomy and Rice Quantum Institute, Rice University, 6100 Main St., Houston TX 77005, USA}
\author{R.~G.~Hulet}
\affiliation{Department of Physics and Astronomy and Rice Quantum Institute, Rice University, 6100 Main St., Houston TX 77005, USA}

\date{\today}

\begin{abstract}

We have studied the effects of a disordered optical potential on the transport and phase coherence of a Bose-Einstein condensate (BEC) of $^7$Li atoms.  At moderate disorder strengths ($V_D$), we observe inhibited transport and damping of dipole excitations, while in time-of-flight images, random but reproducible interference patterns are observed.  \textit{In-situ} images reveal that the appearance of interference is correlated with density modulation, without complete fragmentation.  At higher $V_D$, the interference contrast diminishes as the BEC fragments into multiple pieces with little phase coherence.

\end{abstract}


\maketitle

The behavior of a superfluid/superconductor in the presence of
disorder is of fundamental interest.  A superfluid can flow
without friction around obstacles and a superconductor can have
zero resistance despite material defects.  On the other hand,
disorder is able to localize particles, resulting in an insulating
state \cite{leedis}.  Experimentally, disorder-induced
superfluid/superconductor to insulator transitions (SIT)  have
been probed in many systems, including superfluid helium in porous
media \cite{reppy}, thin film and granular superconductors
\cite{goldman,belo07} and random Josephson junction arrays
\cite{randj}.  While many believe that such a SIT is a quantum
phase transition driven by quantum fluctuations, it remains a
central task to understand exactly how the
superfluid/superconducting order parameter, which consists both an
amplitude and a phase, may be destroyed with increasing disorder.
 Numerous fundamental questions remain, such as the nature of the
insulator, the fate of phase coherence throughout the transition,
and the possibility of intermediate metallic phases
\cite{goldman,fisherdual,pp}.

Cold atoms, with their intrinsic cleanliness coupled with
remarkable controllability of physical parameters, have emerged as
exceptional systems to study various condensed matter problems.
Recently, several experiments \cite{lye,fort,clement,schulte} have
studied $^{87}$Rb condensates in \textit{random} optical potentials, and
observed, for example, damping of collective excitations \cite{lye} and inhibition of expansion
\cite{fort,clement,schulte} due to disorder.  Another experiment
\cite{quasi} has examined a BEC in an incommensurate
(\textit{quasi}-random) optical lattice, in order to investigate a
possible ``Bose-glass" phase \cite{boseg}.  Experiments with
disordered atomic quantum gases may provide unique insights into
disordered quantum systems, and may uncover a rich variety of
quantum phases \cite{disgas}.

Here we report experiments on a BEC of interacting
$^7$Li atoms subject to a well-controlled disordered potential.  While we corroborate previous transport
measurements \cite{lye,fort,clement,schulte}, we have also probed
the ground state density distribution and phase coherence of the disordered BEC
by performing both \textit{in-situ} and time-of-flight (TOF)
imaging.  While disorder inhibits transport of the BEC,
\textit{reproducible} TOF interference patterns are observed for
intermediate disorder strengths $V_D$, reflecting an underlying
phase coherence in the disordered BEC. At stronger $V_D$, the
interference contrast diminishes as the BEC fragments into a
``granular" condensate, which is expected to have no phase
coherence.

We create an optically trapped BEC of $^7$Li atoms in the $m_J$ =
$-$1/2, $m_I$ = 3/2 state at a bias magnetic field $B$ $\simeq$
720 G, using procedures essentially similar to those described in
\cite{strecker}. The bias field tunes the scattering length $a_s$
via a Feshbach resonance, to a large positive value ($\sim$200
$a_0$, where $a_0$ is the Bohr radius), enabling efficient
evaporative cooling \cite{strecker, khay}. The optical dipole trap
is produced by focusing a 1030 nm Yb:YAG laser beam to a Gaussian
1/\textit{e}$^2$ intensity radius of 34 $\mu$m. In the following,
we label the axial direction (along which the optical trap beam
propagates) as $z$, and the two radial directions as $x$ and $y$.
At the final trap depth, the radial trap frequencies $\omega_x$ =
$\omega_y$ $\simeq$ 2$\pi$$\times$180 Hz, and the axial trap
frequency $\omega_z$ $\simeq$ 2$\pi$$\times$3.6 Hz.  Part of the
axial confinement is contributed by a residual curvature from the
bias magnetic field. At this trap depth, the BEC contains
$\sim$5$\times$10$^5$ atoms with no discernible thermal component.
The corresponding chemical potential $\mu$ $\approx$
\textit{h}$\times$1 kHz, where \textit{h} is Planck's constant.

\begin{figure}
\includegraphics[width=8cm]{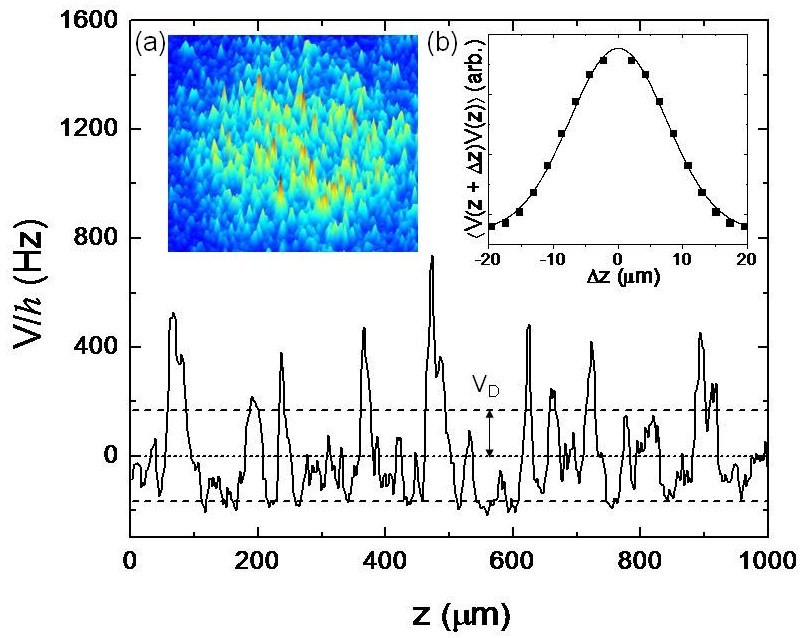}
\caption{\label{fig:speckle} (color) Typical 1D disordered potential
$V(z)$.  (a) False-color surface plot of the disordered speckle
potential $V(y,z)$.  The pixel size corresponds to 2.2 $\mu$m in the plane of the atoms. (b)
Autocorrelation function of $V(z)$.  The line is a fit to
$e^{-2\Delta z^2/\sigma_z^2}$, within a scaling factor.  As expected for fully developed speckle, the intensity follows a negative exponential probability distribution.}
\end{figure}

To create the disordered potential, we pass another beam, derived
from the same laser as used for the optical trap, through a
diffusive plate.  This results in laser speckle \cite{speckle}
similar to that used in \cite{lye,clement}.  This beam propagates along the $x$ direction.  Because of the optical dipole force, the
speckle's spatially varying intensity pattern gives rise to a
disordered potential $V(x,y,z)$ for the atoms.
Figure~\ref{fig:speckle}(a) shows a 2D cross-section, $V(0,y,z)$,
as imaged by a camera, while the main panel shows an axial cut,
$V(z)$ = $V(0,0,z)$, of the disordered potential through the
center ($x$ = $y$ = 0) of the atomic cloud.  The disorder
strength, $V_D$, is proportional to the intensity of the speckle
laser, and is defined as the standard deviation of the mean
of $V(z)$. The disorder correlation length along the $z$
direction, $\sigma_z$, is defined by fitting the autocorrelation
function $\left<V(z)V(z+\Delta z)\right>$ to $e^{-2\Delta z^2/\sigma_z^2}$
(Fig.~\ref{fig:speckle}(b)); we find $\sigma_z$ $\simeq$ 15
$\mu$m, which is much greater than the condensate healing length of $\sim$1 $\mu$m. Since $\sigma_x$ $\gg$ $\sigma_y$ = $\sigma_z$
\cite{speckle,clement} and the radial Thomas-Fermi radius of the
BEC is less than 10 $\mu$m, to a good approximation the BEC is
subject to a 1D disordered potential $V(z)$ = $V(0,0,z)$ along its
axial direction.

After producing a pure (to within our resolution) BEC, the disordered potential is ramped on
in $\sim$100 ms \cite{heatnote} to a specific value of $V_D$,
which we express in terms of $\mu$.  The BEC is then
held in the disordered and trap potentials while various
experiments are performed, as described below.  The final state is
probed by absorption imaging (also along the $x$ direction) of the
atomic cloud released from the disordered and optical trap
potentials, or by \textit{in-situ} phase contrast imaging \cite{bradley}.

\begin{figure}
\includegraphics[width=8cm]{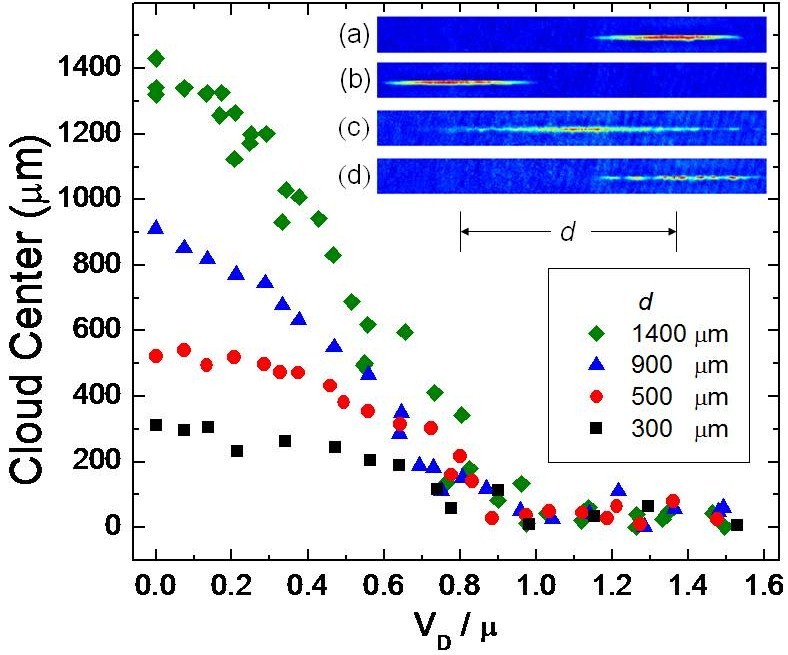}
\caption{\label{fig:drag} (color) Effect of disorder on slow transport of
the BEC.  The trap center is offset to the left for a distance $d$
over a time interval of 700 ms.  (a) Initial state of the BEC,
without disorder or offset.  The center of this initial cloud
defines the reference position ``0".  (b-d) Result of offsetting
the trap by $d$ $\approx$ 1400 $\mu$m, with the cloud subjected to a
disordered potential of various $V_D$: (b) $V_D$ = 0; (c) $V_D$
$\approx$ 0.3 $\mu$; (d) $V_D$ $\approx$ $\mu$.  Images (a-d) are
obtained by absorption imaging taken with a 1 ms TOF after the
cloud is abruptly released from the disordered and optical trap potentials.
Main panel shows the center by weight of the cloud as a function
of $V_D$, for several offsets $d$. $V_D$ and $\mu$ have uncertainties of $\sim$20$\%$.}
\end{figure}

We first describe data characterizing the transport properties of
the disordered BEC. In one type of measurement, we \textit{slowly}
ramp on a magnetic field gradient along the $z$ direction, which
offsets the center of the harmonic trap from $z$ = 0 to $z$ =
$-d$.  The disordered potential is kept stationary during the trap
offset.  As shown in Figs.~\ref{fig:drag}(a) and (b), without
disorder the center of the atomic cloud follows the trap. For
intermediate disorder strength the cloud lags behind the new trap
center and is stretched, as shown in Fig.~\ref{fig:drag}(c). For
stronger disorder, as shown in Fig.~\ref{fig:drag}(d), the cloud
is pinned at its initial position and does not respond to the
offset.  We plot the cloud center as a function of $V_D$ for 4
different offset distances $d$ in Fig.~\ref{fig:drag}.  The cloud
is pinned and does not respond to trap offset for $V_D$ $\agt$
0.8$\mu$, within our experimental time scale and uncertainty.

\begin{figure}
\vspace{0cm}
\includegraphics[width =8.3cm]{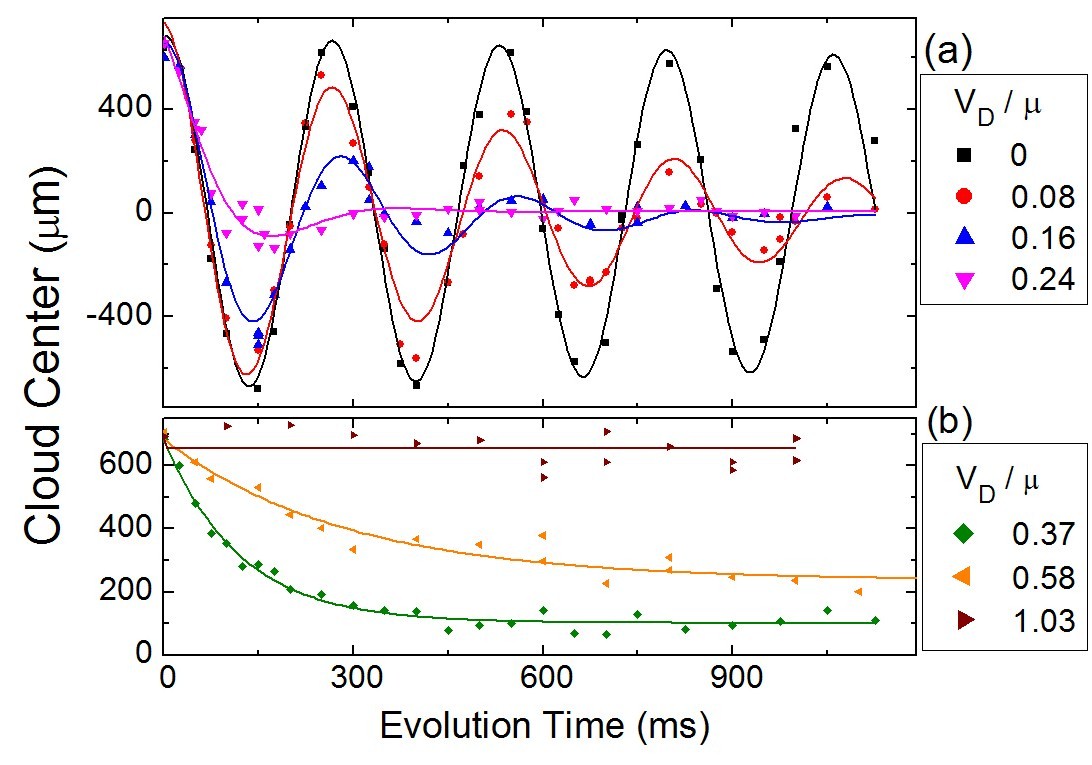}
\vspace{0cm}
\caption{\label{fig:osc} (color) Effect of disorder on dipole excitation of the condensate.  Data shows the cloud center as a function of time following a sudden axial offset of the trap, for various disorder strengths $V_D$ (shown in two separate panels for clarity).  The lines are fits to damped dipole oscillations.}   
\end{figure}

We have also studied the effect of disorder on the dipole
oscillation of the trapped BEC (Fig.~\ref{fig:osc}).  In contrast
to the slow offset experiment discussed above, here the harmonic
trap center is \textit{abruptly} (2 ms) offset, and the cloud
evolves in the presence of the shifted trap and stationary
disorder for a variable time before being released for imaging.
Without disorder, the condensate undergoes undamped dipole oscillations in
the trap with frequency $\omega_z$.  Even disorder as small as
$V_D$ $\sim$ 0.1 $\mu$, however, damps these oscillations.
Increasing the disorder strength to $V_D$ $\agt$ 0.4 $\mu$ causes
the dipole motion to be overdamped, until when $V_D$ $\agt$ $\mu$,
the cloud becomes pinned at its initial position to within our
experimental resolution.

\begin{figure}
\vspace{0cm}
\includegraphics[width =8.4cm]{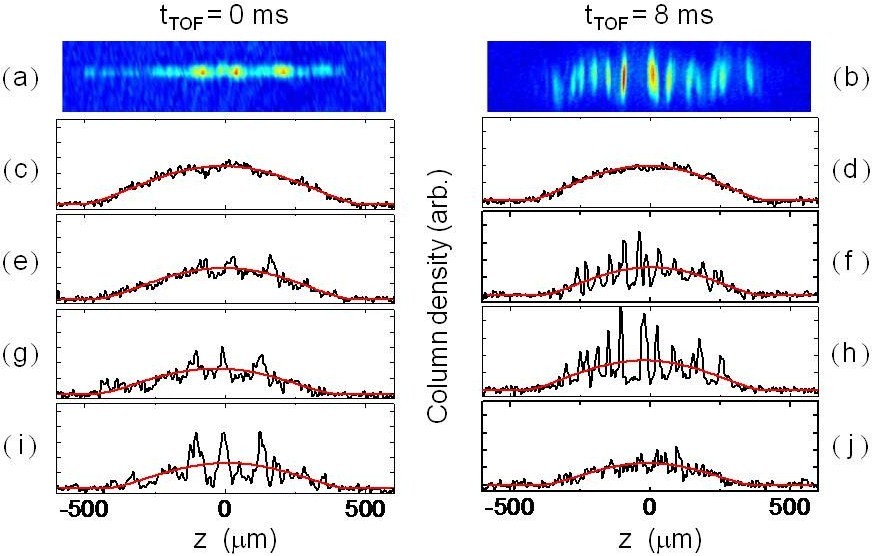}
\vspace{0cm}
\caption{\label{fig:tof} (color) \textit{In-situ} profiles and matter wave
interference.  (a) \textit{In-situ} image for $V_D$ $\approx$ 0.5
$\mu$; (b) TOF image for $V_D$ $\approx$ 0.5 $\mu$.  The TOF 2-D image uses a different color scale from the one used for the \textit{in-situ} 2-D image, and the displayed aspect ratio of the \textit{in-situ} images is reduced by 50\% in order to enhance the transverse size for visibility.  (c-j) \textit{In-situ} column-density profiles on left and corresponding TOF images on right; (c, d) $V_D$ = 0; (e, f) $V_D$ $\approx$ 0.3 $\mu$;
(g, h) $V_D$ $\approx$ 0.5 $\mu$; (i, j) $V_D$ $\approx$ 1.0 $\mu$.  Solid red lines are fits to Thomas-Fermi distributions.  $\omega_z$ $\simeq$ 2$\pi$$\times$2.8 Hz for data shown in Figs.~\ref{fig:tof} and \ref{fig:contrast}.}
\end{figure}

\begin{figure}
\vspace{0cm}
\includegraphics[width = 8cm]{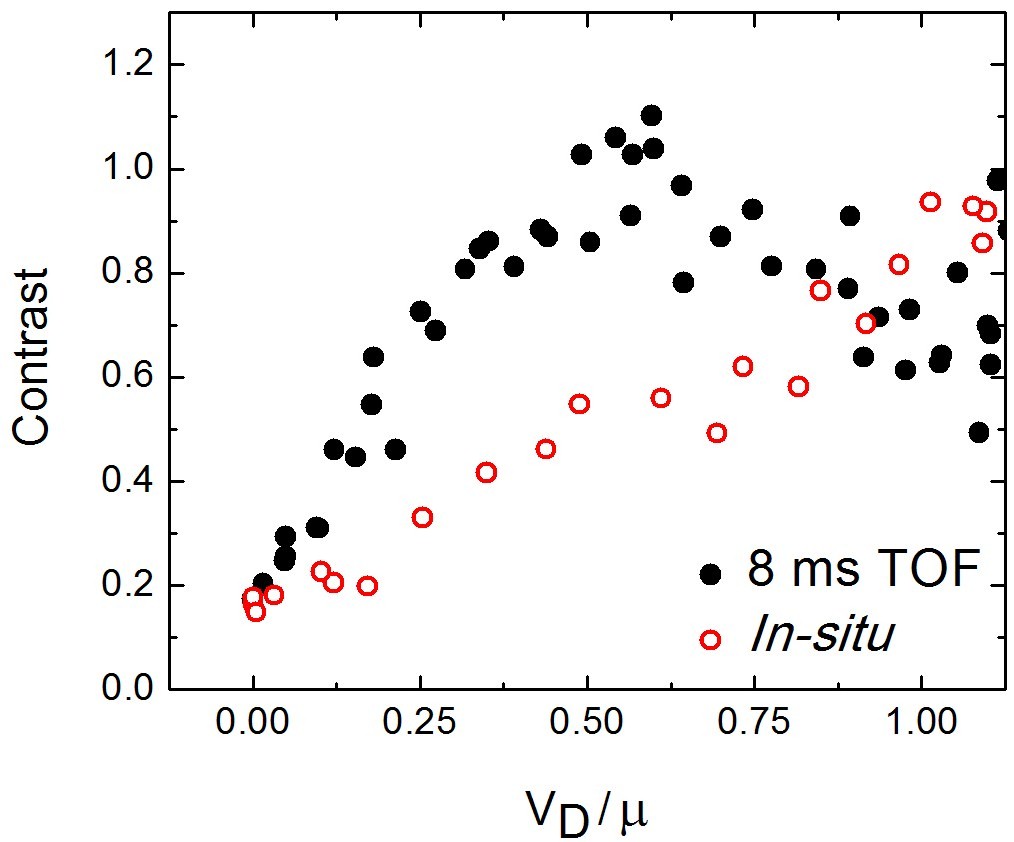}
\caption{\label{fig:contrast} (color online) Contrast of the matter wave
interference pattern in 8 ms TOF images (filled diamonds) and
contrast of \textit{in-situ} density fluctuations (open circles)
as a function of $V_D/\mu$. The contrast is extracted from the
central axial cut of the column density, as its mean deviation from
the Thomas-Fermi fit normalized by the peak value of the fit.}
\end{figure}

In addition to transport measurements, we have performed
\textit{in-situ} \cite{insitu} and TOF imaging on the
disordered BEC, yielding information on both its density and
phase. Figures~\ref{fig:tof}(a) and (b) show \textit{in-situ} and
corresponding TOF images taken 8 ms after releasing the cloud.  For intermediate $V_D$
(Figs.~\ref{fig:tof}(b) and (h)) striking random fringes, which we
interpret as matter wave interference, develop after sufficiently
long TOF expansion of the BEC following release from the optical
potentials.  We note that at such $V_D$, the corresponding
\textit{in-situ} images (Figs.~\ref{fig:tof}(a) and (g)), are
consistent with the cloud still being \textit{connected}.  By
increasing $V_D$ above $\mu$, the disordered BEC becomes
\textit{fragmented} shown in the \textit{in-situ}
profile, Fig.~\ref{fig:tof}(i), and the fringe contrast observed
in TOF diminishes, as shown in Fig.~\ref{fig:tof}(j). To
quantify the dependence of the TOF fringe pattern on $V_D$, we
plot both the TOF and \textit{in-situ} contrast vs. $V_D/\mu$ in
Fig.~\ref{fig:contrast}. The TOF fringe contrast peaks at $V_D$
$\simeq$ 0.5 $\mu$, while the contrast in the \textit{in-situ} images increases up to the highest $V_D$ investigated.


Similar, but less well resolved fringes in TOF images have been
reported previously \cite{lye,schulte}.  Our highly elongated BEC
facilitates a systematic and quantitative study of the fringes. We
find the positions of these irregularly spaced fringes are
\textit{reproducible} in repeated measurements, even with
different holding times of the BEC in the disordered potential.
During preparation of this manuscript, we became aware of a
closely related paper \cite{preprint}, which also reports
reproducible fringes in TOF images of a disordered BEC. The
reproducibility suggests that these fringes are unlikely due to
some initial (before release) phase fluctuations \cite{dettmer} in
the disordered BEC, or due to interference of a \textit{few} BECs
separated by high potential barriers with no well defined relative
phase between them, as in either of these scenarios the fringe
positions are not expected to be reproducible.

Numerical simulations based on the Gross-Pitaevskii equation \cite{schulte, preprint} have shown that
such interference can occur when moderate disorder-induced density
fluctuations cause different parts of a \textit{phase-coherent}
BEC to expand with different velocities and overlap. As $V_D$ is
increased from 0, the contrast of the interference is expected to
increase, consistent with the data shown in Fig.~\ref{fig:contrast}
at small $V_D$.  When $V_D$ becomes sufficiently large, however,
the BEC fragments into multiple pieces, as is seen
in the \textit{in-situ} image, Fig.~\ref{fig:tof}(i).
An array of \textit{randomly spaced} condensates should not produce visible
interference in TOF \cite{fallani,zoran}, again consistent with our observation of
diminishing fringe contrast at high $V_D$ (Figs.~\ref{fig:tof}(j)
and ~\ref{fig:contrast}).


At high $V_D$, we observe completely inhibited transport,
indicating a transition to an insulator.  This was seen in
previous transport experiments, where the nature of the insulator
is inferred to be that of a fragmented BEC
\cite{clement,fort,lye}.  Our \textit{in-situ} images give the
first direct observation of fragmentation of a highly disordered
BEC. Because of the exponentially suppressed Josephson tunneling
between fragments due to high potential barriers, the fragmented
BEC is expected to have large phase fluctuations and no phase
coherence \cite{laurent}. This regime is analogous to a granular
superconductor \cite{goldman,belo07}, which is an insulator
without global phase coherence.




Dipole oscillations are damped at moderate $V_D$, as shown in
Fig.~\ref{fig:osc}, and in agreement with previous observations
\cite{lye}. Excitations such as solitons, vortices and phase slips
have been suggested \cite{diss} as possible mechanisms giving rise
to damping. Our \textit{in-situ} as well as TOF data at moderate
$V_D$ suggest a disordered BEC that is still connected (not yet
fragmented) and phase coherent, in order to give a reproducible
interference pattern.  Such a state may provide insight into the
physics of ``homogenous" disordered superconducting films,  where
there has been tremendous interest regarding the form of the order
parameter, possibilities of metallic states, and the fate of phase
coherence and phase stiffness \cite{andersonb} during the SIT
\cite{goldman,pp}.  Damped dipole oscillations were also recently
studied for a BEC in a periodic potential \cite{demarco} in the
context of a possible ``Bose metal" phase \cite{pp}.

Strong repulsive interactions are
believed to preclude Anderson localization
(AL) \cite{schulte, laurent} in a BEC such as ours and those in previous
experiments where the condensate healing length is small in comparison to $\sigma_z$ \cite{lye,fort,clement,schulte}. Interactions,
however, can be significantly reduced in $^7$Li due to an
extremely gradual zero-crossing in the $B$ dependence of $a_s$
\cite{strecker,khay}, creating more favorable conditions to
observe AL.  Such experiments are currently underway.  A similar
zero-crossing has been identified in $^{39}$K \cite{roati}.







Financial support of this work was provided by NSF, ONR, and the Welch (C-1133) and Keck Foundations. Y.~P.~C. also acknowledges the support of a J.~Evans Attwell-Welch Postdoctoral Fellowship. We thank A.~Aspect, L.~Sanchez-Palencia, and T.~A.~Corcovilos for helpful discussions.


\begin{thebibliography}{}
\bibitem{heatnote} We have checked that ramping on of the disordered potential does not produce significant heating by ramping the disordered potential back to zero to recover a pure condensate.
\end{thebibliography}
\end{document}